\documentstyle[12pt]{article}

\newcommand{\eq}[1]{eq.(\ref{#1})}

\newcommand{\vpar}[2]{\frac{\delta #1}{\delta #2}}

\def\ltap{\raisebox{-.55ex}{\rlap{$\sim$}} \raisebox{.4ex}{$<$}}
\def\gtap{\raisebox{-.55ex}{\rlap{$\sim$}} \raisebox{.4ex}{$>$}}

\def\e{\mbox{e}}

\begin{document}
\title{Instanton-like transitions at high energies in (1+1) dimensional
scalar models. II. Classically allowed induced vacuum decay.}
\author{
  V.~A.~Rubakov and D.~T.~Son\\
  {\small \em Institute for Nuclear Research of the Russian Academy of
  Sciences,}\\
  {\small \em 60th October Anniversary prospect 7a, Moscow 117312}\\
  }
\date{January 1994}

\maketitle
\begin{abstract}
  We consider classical Minkowskian solutions
 to the field equation in the (1+1)
dimensional  scalar model with the exponential interaction
that describe the unsuppressed false vacuum decay induced by $n$
initial particles. We find that there is a critical value of $n$ below
which there are no such solutions, i.e.,
 the vacuum decay is always suppressed. For the number of initial
particles larger than this value the vacuum decay is unsuppressed at high
enough energies.
\end{abstract}

\newpage

\section{Introduction}

Instanton--like processes induced by collisions of
highly energetic particles remain an
object of controversial discussion (for reviews see, e.g.,
\cite{Mrev,Trev}). One particular type of these processes is the induced
decay of the false vacuum in scalar theories
\cite{Affleck,VolSel,Vol,RSTind,Kisel,GorVol,Rutgers}. Being exponentially
suppressed at zero energy by the action of a bounce \cite{Coleman,CalCol},
the rate exponentially increases with the energy of incoming particles,
 and may or may not become large at high enough energies.
Similar behavior is inherent in the baryon number violating transitions in
the electroweak theory and analogous processes in other models with
instantons \cite{Ringwald,Espinosa}.

Although the two--particle initial states are of primary interest in this
problem, it has been argued that it is instructive to study the
multiparticle initial states and then consider the limit of small number of
initial particles \cite{RT,T,RST,Mueller}. If the
number of initial particles
is of order $1/g^2$ , where $g$ is small dimensionless coupling constant,
the induced instantion--like transition rate is calculable, at least
 in principle, in a semiclassical way. The corresponding classical
 solution is smooth on a particular contour in complex time plane and
 is determined by boundary conditions specifying the number of particles,
 as discussed in ref.\cite{RST}.

The classical boundary value problem relevant to the false vacuum decay
induced by multiparticle ''collisions'' has been studied in
our previous paper \cite{Rutgers} in
(1+1) dimensional theory with the Lagrangian
\begin{equation}
  L = {1\over 2}(\partial_{\mu}\phi)^2 - V(\phi),
  \label{Lagrangian}
\end{equation}
where the potential $V(\phi)$ has the following form,
\[
  V(\phi) =    {m^2\over 2}\phi^2 -
      {m^2v^2 \over 2}\exp\left[2\lambda\left({\phi\over v}-1\right)\right].
\]
 The dimensionless parameter $v$ plays the role of the inverse coupling
constant, while $\lambda$ is an additional dimensionless parameter of the
theory. For the semiclassical calculations to be reliable, one
requires $v\gg\lambda$, while the actual classical solutions can be
explicitly found at $\lambda\gg 1$. Thus, the convenient regime is
\[
  v\gg\lambda\gg 1.
\]
In this regime, the potential $V(\phi)$
is quadratic at $\phi<v$ and has steep cliff at
$\phi>v$, as shown in fig.1;
the false vacuum is $\phi=0$, while the true vacuum is $\phi=+\infty$.

When the energy of the initial state $E$ is roughly of order $mv^2$ (but
not too high), and the number of incoming particles is of order $v^2$, the
rate of the induced decay is suppressed by a semiclassical factor
\[
  \Gamma(E,n)\propto\e^{-F(E,n)}
\]
where
\[
  F(E,n)=v^2\tilde{F}\left({E\over E_{sph}},{n\over n_{sph}},\lambda\right)
\]
and
\[
  E_{sph}=mv^2
\]
\[
  n_{sph}={2\over\pi}v^2
\]
are the energy and number of particles for the sphaleron (critical bubble)
in this model.

At $n/n_{sph}\gg\lambda^{-1}$, the suppression factor has been calculated
\cite{Rutgers} in a wide interval of energies. It has been argued that
there exists some critical energy $E_{crit}(n)$ at which $F$ becomes equal
to zero, i.e., the exponential suppression disappears. The dependence
of the critial energy on the number of initial particles is particularly
simple at $1\gg n/n_{sph}\gg\lambda^{-1}$,
\begin{equation}
  E_{crit}={4\over\pi^3}\exp\left({\pi^2\over 4}{n_{sph}\over n}-1\right)
           \left({n\over n_{sph}}\right)^2E_{sph}
\label{3*}
\end{equation}
Notice that $E_{crit}$ is much larger than the sphaleron energy.

It remained unclear whether the exponential suppression of the induced
vacuum decay is exponentially suppressed or not at any energies for even
smaller number of incoming particles, $n/n_{sph}~\ltap~\lambda^{-1}$.
For these $n$, it has
been found only that  the exponent $F$ is substantially
different from its vacuum value  at exponentially high energies,
\begin{equation}
  \ln{E\over E_{sph}}\sim\lambda
\label{3**}
\end{equation}
Eq.(\ref{3**}) sets roughly the energy scale of interest in the case
$n/n_{sph}~\ltap~\lambda^{-1}$.

Coming back to the case $n/n_{sph}\gg\lambda^{-1}$, the absence
 of the exponential suppression at $E>E_{crit}(n)$ signalizes
 that there
exist {\em classical} transitions from an initial state with $n$ particles
above the false vacuum, to the true vacuum. These transitions are described
by real Minkowskian solutions to the field equations
(the real-valuedness of the solutions at large time is ensured by
the summation over all final states \cite{RST}; then the solutions
are obviously real in the whole Minkowski space-time). At $t\to-\infty$,
these solutions should be the collections of plane waves above
the false vacuum, $\phi=0$,
\[
  \phi(x,t)=\int{dk\over 2\pi}\left[\e^{ikx-i\omega_kt}f_k+
            \e^{-ikx+i\omega_kt}f^*_k\right]
\]
They correspond to initial coherent states with
the energy and number of incoming particles
\[
  E={1\over\pi}\int dk\omega_k^2f^*_kf_k
\]
\[
  n={1\over\pi}\int dk\omega_kf^*_kf_k
\]
For this correspondence between the classical solutions and quantum
 states be exact, one requires that $E$ and $n$ are of order $v^2$,
up to arbitrary dependence on $\lambda$. It is this regime that will
be considered throughout this paper.
To describe the induced decay of the
 false vacuum, the solution should have a
singularity $\phi\to+\infty$ at finite time, because the (homogeneous)
field rolling down the cliff of $V(\phi)$ reaches
the "true vacuum" $\phi=\infty$ in finite
time.

At large enough, but fixed energy and number of initial particles,
there may exist a variety of initial coherent states that
induce the false vacuum decay without suppression.
 In other words, there may exist a variety
of the corresponding classical solutions. So, finding these solutions
 is not a well formulated mathematical problem, unlike the boundary
value problem, with expectedly unique solution at given $E$ and $n$,
relevant to the suppressed induced vacuum decay at relatively low
 energies and number of incoming particles. In reality this means that
 one searches for (Minkowskian) solutions by making use of a certain
Ansatz. The existence of a solution at given $E$ and $n$ within this
Ansatz guarantees that the false vacuum decay is classically allowed
at these $E$ and $n$, but the absence of such a solution within the
Ansatz tells essentially nothing. We will present one useful Ansatz
in the model (\ref{Lagrangian}), as well as the corresponding set of
explicit solutions, in sect.2.2. We will see that our solutions exist at
$n>\pi^2n_{sph}/\lambda$ and high enough energies, and, furthermore,
these solutions at $1\gg n/n_{sph}\gg\lambda^{-1}$ appear for
the first time at energies that coincide, up to a pre-exponential
numerical factor, with the critical energy, eq.(\ref{3*}). The latter
 result, combined with our previous results \cite{Rutgers}, gives a
 coherent picture of the induced vacuum decay at
 $1\gg n/n_{sph}\gg\lambda^{-1}$: the decay rate
is exponentially suppressed
at $E<E_{crit}(n)$, grows towards $O(1)$ as $E$ increases
 towards $E_{crit}(n)$,  and the decay
proceeds classically at $E>E_{crit}(n)$.

The absence, within the Ansatz of sect.2.2, of the relevant Minkowskian
solutions at $n<\pi^2n_{sph}/\lambda$ may be at best viewed as a loose
indication that the induced vacuum decay is classically forbidden
 at all energies for such $n$. To see that this is indeed the case,
one has to find arguments independent of any particular Ansatz. We present
two different arguments in sects.3 and 4; although we beleive that
each of these arguments is sufficient by itself, we discuss both of
them to illustrate slightly different techniques that may be of use in more
complicated models.

In sect.3 we show that at any energy, no real Minkowskian solutions
describing the induced vacuum decay exist at $n<n_{crit}$ where
 $n_{crit}=\pi^2n_{sph}/\lambda$.

In sect.4 we consider, instead of the number of incoming particles,
another operator,
\[
\hat{A}=\int dk \frac{1}{\omega_{k}}a^{\dagger}_ka_k
\]
to label initial states. The limit of a few highly energetic particles
corresponds to the limit $A/v^{2}\to 0$, while we show in sect.4 that
at given energy $E$, the classical solutions describing the induced vacuum
decay exist only at $A/v^{2}>B(E)$, where $B(E)$ is strictly positive
at any $E$. This again means that the vacuum decay induced by collisions
of a small number of particles is exponentially suppressed at all
energies.

Sect.5 contains concluding remarks.

\section{Classical scattering at high energies}

\subsection{Construction of classical solutions}

The Minkowskian field equation for the model with the Lagrangian
(\ref{Lagrangian}) has the following form,
\begin{equation}
  \partial_{\mu}^2\phi=-m^2\phi+\lambda m^2v\exp\left[2\lambda\left(
                      {\phi\over v}-1\right)\right]
  \label{fieldeq}
\end{equation}
The general solution to eq.(\ref{fieldeq}) is unknown.
 However, at large
$\lambda$ and high enough energy per an incoming particle,
$E/n\gg m$, the solutions of interest can be found explicitly.
Indeed, the incoming particles are ultra-relativistic, so one can
neglect their masses near the interaction region where the field
is non-linear. Due to the large value of $\lambda$, the size of the
interaction region, $r_{0}$, is small, $r_{0}\ll m^{-1}$.
 Inside the interaction region
(i.e. at $x$, $t~\ltap~r_0$), the mass term in eq.(\ref{fieldeq}) can be
neglected, so the field equation is reduced to the Liouville equation,
\begin{equation}
  \partial_{\mu}^2\phi=\lambda m^2v\exp\left[2\lambda\left(
                      {\phi\over v}-1\right)\right],
  \label{Liouveq}
\end{equation}
whose solutions are known explicitly.
 In contrast, at large distances, i.e.
at $x$, $t~\gtap~m^{-1}$, only the mass term is essential on the right
hand side of
eq.(\ref{fieldeq}), so the field is  a solution to the massive
Klein--Gordon equation. At intermediate distances, both the mass term and
the exponential interaction term in eq.(\ref{fieldeq}) are inessential, so
$\phi$ obeys the free massless field equation, $\partial_{\mu}^2\phi=0$.
The way to find a solution to eq.(\ref{fieldeq})
 is to solve it
separately in the two regions of small and large distances, $x$,
$t~\ltap~r_0$ and\\
 $x$, $t~\gtap~m^{-1}$, and then match the two field
configurations at intermediate values\footnote{This
approach, with appropriate modifications, has been used in
 ref.\cite{Rutgers} to find interesting solutions to euclidean
field equations in our model; in particular, the instanton (bounce)
and sphaleron (critical bubble) have been obtained in this way to the
leading order in $\lambda$.}\\ of $x$, $t$.

Let us first write down the general solution to the Liouville equation,
\begin{equation}
  \phi(x,t)={v\over 2\lambda}\ln{f'(x+t)g'(x-t)\over\left[1+{1\over a^2}
            f(x+t)g(x-t)\right]^2}
  \label{Liouvsol}
\end{equation}
where $f$ and $g$ are arbitrary real functions of the light "cone"
coordinates $x+t$ and $x-t$, respectively, $f'$ and $g'$ denote derivatives
of $f$ and $g$ with respect to their variables, and
\[
a={2\e^{\lambda}\over\lambda m}
\]
Note that the right
hand side of eq.(\ref{Liouvsol}) is invariant under the
following transformation,
\[
  f\to {Af+B\over Cf+D}
\]
\[
  g\to a^2{Dg-Ca^2\over Aa^2-Bg}
\]
where $A$, $B$, $C$, $D$ are arbitrary real constants.
 We make use of
this ambiguity to impose the following conditions on $f$ and $g$,
\begin{equation}
  f(-\infty)=g(+\infty)=0.
  \label{limit}
\end{equation}
Finally, we assume that $g$ and $f$ are related in a simple way,
\begin{equation}
  g(z)=-f(-z),
\label{5*}
\end{equation}
so the field configuration $\phi(x,t)$ is invariant under spatial reflection
$x\to-x$, $\phi(x,t)=\phi(-x,t)$. In other words, we consider the
classical scattering of two identical, up to the reflection, initial
wave packets; this ensures, in particular, that the total
spatial momentum is
zero.

{}From eq.(\ref{Liouvsol}) it is clear that the final state of the field
configuration possesses the singularity,
$\phi=+\infty$, i.e., describes the transition to the true vacuum,
 provided there exist such $x$ and
$t$  that
\[
  f(x+t)g(x-t)=-a^2.
\]
Clearly, it is sufficient to require that the singlarity exists at $x=0$
at  some $t$, which means that we impose the
condition that,
\[
  f(t)g(-t)=-a^2.
\]
i.e.,
\begin{equation}
  f^{2}(t)=a^{2}
\label{6*}
\end{equation}
at some $t$.

Let us now match this solution to free massive field at negative $t$.
 Taking the formal limit $t\to-\infty$ in
eq.(\ref{Liouvsol}), and making use of the fact that the product\\
 $f(x+t)g(x-t)$ tends
to zero at all  $x$ due to eq.(\ref{limit}), one finds
that $\phi$ is a sum of two free waves moving along the light ``cones'',
\begin{equation}
  \phi(x,t)={v\over 2\lambda}(\ln f'(x+t)+\ln g'(x-t))
  \label{asympt}
\end{equation}
 This field configuration should be  matched to the massive linear
tail. Generally speaking,  one
 represents $\phi$ by the sum of plane waves,
\begin{equation}
  \phi(x,t)=\int dk\left(f_k\e^{ikx-i|k|t}+f^*_k\e^{-ikx+i|k|t}\right)
  \label{massless}
\end{equation}
The massive field configuration that is smoothly matched to the solution
to the Liouville equation can be obtained by replacing
$|k|$ by $\omega_k = (k^2 + m^2)^{1/2}$ in eq.(\ref{massless}),
\begin{equation}
  \phi(x,t)=\int dk\left(f_k\e^{ikx-i\omega_kt}+
            f^*_k\e^{-ikx+i\omega_kt}\right)
  \label{massive}
\end{equation}
In this way one obtains the field configuration both in the region of
non-linearity and in the asymptotics $t\to -\infty$. The conditions
(\ref{limit}) ensure that the field
at large negative time indeed describes linear excitations
above the false vacuum.

This procedure of obtaining the solutions to the field equations
is fairly general. The most non-trivial part of the problem is to
find a suitable function of one real argument, $f(z)$, which obeys
eqs.(\ref{limit}) and (\ref{6*}) and corresponds to a given number of
particles and energy.

\subsection{Examples}

Let us illustrate the above technique by the following choice of
the set of
functions $f$,
\begin{equation}
  f(z)=\int\limits_{-\infty}^{z}dz'{C\over(z'^2+\tau^2)^{\eta}}
  \label{ansatz}
\end{equation}
where $\eta$ and $\tau$ are real parameters and the constant
 $C$  will be defined below.
 At large negative time, eq.(\ref{asympt}) gives
\[
  \phi(x,t)={v\over 2\lambda}\left[-\eta\ln(x^2-(t-i\tau)^2)
            -\eta\ln(x^2-(t+i\tau)^2)+2\ln C\right]
\]
The solution to the massive Klein--Gordon equation that is
smoothly matched to this configuration is
\[
  \phi(x,t)={v\eta\over 2\lambda}\left[ K_0(m\sqrt{x^2-(t-i\tau)^2})
            + K_0(m\sqrt{x^2-(t+i\tau)^2})\right]
\]
provided the value of the constant $C$ is
\[
  C=\left({m\e^{\gamma}\over 2}\right)^{-2\eta}
\]
where $\gamma$ is the Euler constant. For the existence of singularities
ensuring the formation of the true vacuum, one requires (see eq.(\ref{6*}))
\[
  f(+\infty)=\int\limits^{+\infty}_{-\infty}dz'{C\over (z'^2+\tau^2)^{\eta}}
             > a = {2\e^{\lambda}\over\lambda m}
\]
This relation gives the following upper bound on the value of $\tau$,
\begin{equation}
  \tau < \tau_{crit} = \e^{-\lambda/2\eta-1}{2\over m\e^{\gamma}}
     \left[{\sqrt{\pi}\Gamma(\eta-1/2)\over\e^{\gamma}\Gamma(\eta)}\right]
     ^{1/(2\eta-1)}
  \label{taucrit}
\end{equation}

For calculating the energy and number of initial particles, one
evaluates the
Fourier components of $\phi$ in the initial asymptotics,
\[
  \phi(k) = {\eta v\over\lambda}{\pi\over\omega_k}\e^{-\omega_k\tau}
\]
The case of small enough $\tau$, such that
\[
  \ln{1\over m\tau}\gg 1
\]
 will be of special interest (note that this relation is valid for
$\tau \sim \tau_{crit}$). The number of particles and their energy
are then
\begin{equation}
  n = {1\over\pi}\int\limits^{+\infty}_{-\infty}\omega_k|\phi(k)|^2
    =  {2\pi\eta^2v^2\over\lambda^2}\left[\ln{1\over m\tau}
       -\gamma+\ln 2\right]
  \label{n}
\end{equation}
\begin{equation}
  E = {1\over\pi}\int\limits^{+\infty}_{-\infty}\omega_k^2|\phi(k)|^2
    =  {\pi\eta^2v^2\over\lambda^2}{1\over\tau}
  \label{E}
\end{equation}
Thus, the energy and number of incoming particles are expressed
through the two parameters of our solution, $\eta$ and $\tau$.
Alternatively, $\eta$ and $\tau$ may be viewed as functions of $E$
and $n$. Eq.(\ref{taucrit}) required for the solution to describe the
false vacuum decay, and not merely classical scattering above the
false vacuum, becomes then the constraint on $E$ and $n$, which we
now discuss in some detail.

The curve
\begin{equation}
  \tau(E,n) = \tau_{crit}(E,n)
\label{V9*}
\end{equation}
divides the $(E,n)$ plane into two parts, as shown in fig.2
($\tau_{crit}$ depends on $E$ and $n$ through $\eta(E,n)$). Above
this line (region II above the dashed line) eq.(\ref{taucrit}) is
valid, and our solution indeed describes the induced vacuum decay.

Let us consider first the case $n/n_{sph}\gg 1/\lambda$. It follows
from eqs.(\ref{n}) and (\ref{E}) that this case corresponds to
$\eta \gg 1$ (roughly speaking,
$\eta=O(\lambda)$).
Eqs.(\ref{n}),(\ref{E}) and (\ref{V9*}) determine the curve
parametrically, $\eta$ being the parameter along the curve,
\begin{equation}
  n = {\pi^2\eta\over 2\lambda}n_{sph}
\label{7*}
\end{equation}
\begin{equation}
  E = E_{sph}{\pi\eta^2\e^{\gamma}\over 2\lambda^2}
      \exp\left({\lambda\over 2\eta-1}\right)
\label{8*}
\end{equation}
It is clear from these equations that the critical energy at which
scattering leads to the vacuum decay increases as the number of
particles becomes smaller. The form of this curve is particularly
simple when $1\gg\ n/n_{sph}\gg \lambda^{-1}$: our
configuration describes vacuum decay when
\[
  E > E_{crit}(n) = {2\e^{\gamma}\over\pi^3}\left({n\over n_{sph}}\right)^2
                 \exp\left({\pi^2\over 4}{n_{sph}\over n}\right)E_{sph}
\]
Surprisingly enough, for a given number of initial particles $n$, the
minimal energy when the configuration describes the vacuum decay,
$E_{crit}(n)$, is larger than the one found in ref.\cite{Rutgers},
eq.(\ref{3*}), only by
a numerical factor  $e^{1+\gamma}/2$.

As discussed in sect.1, the critical energy found within the particular
Ansatz, eq.(\ref{ansatz}), might have been arbitrarily different from
the true critical energy. However, our solution gives the existence
proof that the induced vacuum decay is classically allowed at
$n/n_{sph}\gg\lambda^{-1}$ and sufficiently high energy.

Let us now consider our solution at $n/n_{sph}\sim \lambda^{-1}$.
This regime occurs when
 $\eta=O(1)$, and the critical curve has the following parametric form,
\[
  {n \over n_{sph}} = {\pi^2 \over \lambda} {\eta^2 \over 2\eta-1}
\]
\[
  {E \over E_{sph}} \sim \exp\left({\lambda \over 2\eta-1}\right)
\]
Note that in this case  $E$ is
exponential in $\lambda$, see fig.2.
Two properties of this curve are of particular interest. First, the
minimum value of $n$ is
\begin{equation}
n_{crit}={\pi^2\over\lambda}n_{sph}
\label{V13*}
\end{equation}
This means that the induced vacuum decay does not proceed
classically at any energy,
 within our Ansatz, at $n<n_{crit}$. We will see in
sect.3 that this property holds for all classical solutions, i.e.,
the vacuum decay is classically forbidden at any energy
 (independently of any
Ansatz) at  $n<n_{crit}$, where
  $n_{crit}$ is given by eq.(\ref{V13*}).

Second, at $\ln E/E_{sph}>\lambda$, the number of particles $n$ at
the curve increases with energy (dashed line in fig.2). This means
that at fixed $n$ greater than $n_{crit}$, our solution describes the
vacuum decay only in a finite interval of energies, i.e., the vacuum
decay appears  to be again suppressed at high energies. This
unexpected behavior is actually a peculiarity of our Ansatz: it is
clear that increasing the energy at fixed number of particles cannot
make the vacuum decay exponentially suppressed once it is classically
allowed at lower energies. Indeed, the initial state can spend some
fraction of its energy to radiate perturbatively a few highly
energetic particles, and the bulk of remaining particles would
classically induce the vacuum decay at effectively lower energy.
 Alternatively, the energy can be distributed among the initial
particles in such a way
 that  very few particles would carry most of the  energy while
the others have just the right amount of energy.
At the level of classical solutions the latter possibility
is realized by a trivial generalization of the Ansatz (\ref{ansatz}),
\[
  f(z) = \int\limits_{-\infty}^{z}dz'{C\over(z'^2+\tau_1^2)^{\eta_1}
         (z'^2+\tau_2^2)^{\eta_2}}
\]
where one assumes the following relations between the parameters $\tau_1$,
$\tau_2$, $\eta_1$, $\eta_2$,
\[
  \tau_1\ll\tau_2,~~~\eta_1\ll\eta_2
\]
\[
  \eta_1^2\ln{1\over m\tau_1} \ll \eta_2^2\ln{1\over m\tau_2},
\]
\[
  {\eta_1^2\over\tau_1} \gg {\eta_2^2\over\tau_2}.
\]
The singularity in the
 function $f'$ at $z=\pm i\tau_1$ corresponds to a few highly
energetic particles, while the other singularity, $z=\pm i\tau_2$,
 corresponds to
larger number of particles, each with much lower energy. The number of
particles and the energy of the initial state can be sraightforwardly
 calculated.
The number of particles depends on $\tau_2$ and $\eta_2$ only,
\[
  n = {2\pi v^2\eta_2^2\over\lambda^2}\ln{1\over m\tau_2}
\]
i.e. comes entirely from the soft component, while the expression for the
energy contains $\tau_1$ and $\eta_1$ only,
\[
  E = {\pi v^2\over\lambda^2}{\eta_1^2\over\tau_1}
\]
The condition $f(\infty)>a$, as one can easily see, puts a
restriction on $\tau_2$ and
$\eta_2$ only. Since there is no direct relationship between the energy and
the number of particles, these solutions describe the unsuppressed
induced vacuum decay at arbitrarily high energy for a given
number of particles (if $n>n_{crit}$). So, there is no suppression for
processes with $n>n_{crit}$ at very high energies and the actual
curve dividing the regions of the induced vacuum decay and classical
scattering in the false vacuum is the one shown in fig.2 by the solid
line.

\section{Minimum  number of incoming particles needed for inducing
vacuum decay}

The technique of sect.2.1 enables us to evaluate the minimum number
of incoming particles required for inducing the classically allowed
vacuum decay, no matter how high the energy is. Let us recall that right
before the collision,
the field is a sum of two wave packets moving along the two
light ``cones'',
\[
  \phi(x,t) = \phi_L(x+t) + \phi_R(x-t).
\]
where
\begin{equation}
  \phi_L(z) = {v\over 2\lambda}\ln f'(z),~~~
  \phi_R(z) = {v\over 2\lambda}\ln g'(z).
\label{10*}
\end{equation}
We consider parity symmetric initial states, eq.(\ref{5*}), so the
number of incoming particles is
\begin{equation}
  n = n_L + n_R =
       {2\over\pi}\int\limits_0^{\infty}dk\omega_k\phi_L(k)\phi_L(-k)
  \label{nphi}
\end{equation}
where
\[
  \phi_L(k) = \int dz\phi_L(z)\e^{-ikz}
\]
Our purpose is to minimize $n$ over all possible choices of $\phi_L$ (or
$f$). However, not all functions $f$ are possible: for the configuration to
describe the vacuum decay, one  requires that
\begin{equation}
  f(\infty) = \int\limits_{-\infty}^{\infty} dz f'(z)
            = \int\limits_{-\infty}^{\infty} dz\exp\left({2\lambda\over v}
              \phi_L(z)\right) > a
  \label{finf}
\end{equation}
Indeed, the incoming field (\ref{10*}) is non-singular only if
$f'(z)>0$ for any $z$. Therefore, if $f(\infty)<a$, then the basic
relation, eq.(\ref{6*}) is not valid at any time, and the solution
describes scattering above the false vacuum that does not lead to the
transition to the true one. Let us see that eq.(\ref{finf}) imposes
the energy--independent constraint
\begin{equation}
  n<n_{crit}
\label{V16*}
\end{equation}
where $n_{crit}$ is given by eq.(\ref{V13*}).

To this end, let us discuss slightly more general problem. Namely,
let us  find the minimum of $n$ for a given and fixed value of
$f(\infty)$. Introducing the Lagrange multiplier $\Lambda$, one writes
\begin{equation}
  \vpar{}{\phi_L(-k)}(n-\Lambda f(\infty)) = 0
  \label{condminn}
\end{equation}
Substuting eqs.(\ref{nphi}) and (\ref{finf}) into eq.(\ref{condminn}), one
obtains
\begin{equation}
  \omega_k\phi(k) = {\pi\lambda\over v}\Lambda\int dz
                    \exp\left({2\lambda\over v}\phi_L(z)\right)\e^{-ikz}
  \label{eqminn}
\end{equation}
At high energy per particle, $E/n\gg m$, the frequency
 $\omega_k$ may be replaced by $|k|$
in this equation.

Eq.(\ref{eqminn}) optimizes the form of the incoming wave packet
$\phi_L(z)$ at given $f(\infty)$ in such a way
 that the number of particles
is minimal.
Let us check that  the
 solution to
 eq.(\ref{eqminn}) is given by the function $f(z)$ considered in sect.2.2,
eq.(\ref{ansatz}), with $\eta=1$.
At $\eta=1$ eq.(\ref{ansatz}) reads, in terms of $\phi_L$,
\[
  \phi_L(z) = -{v\over 2\lambda}\ln\left[{m^2\e^{2\gamma}\over 4}
               (z^2+\tau^2)\right]
\]
One has
\[
  \phi_L(k) = \int dz\phi_L(z)\e^{-ikz} =
 \frac{\pi v}{\lambda |k|}\e^{-|k|\tau}
\]
\[
  \int dz\exp\left({2\lambda\over v}\phi_L(z)\right)\e^{-ikz} =
  {4\pi\over m^2\e^{2\gamma}}{1\over\tau}\e^{-|k|\tau}
\]
It is now clear that  eq.(\ref{eqminn}) is satisfied by our
ansatz if $\Lambda$ is related to $\tau$ as follows,
\[
  \Lambda = {v\over\pi\lambda^2}{m^2\e^{2\gamma}\tau\over 4}
\]
So, we have found the solution with the
 minimal number of particles among
all solutions with fixed $f(\infty)$.
The value of $f(\infty)$ for this configuration is
\begin{equation}
  f(\infty)=\frac{4\pi}{m^{2}\e^{2\gamma}}\frac{1}{\tau}
\label{V18*}
\end{equation}
This relates the only parameter of the solution, $\tau$, to
$f(\infty)$. The number of incoming particles is given by eq.(\ref{n})
with $\eta = 1$ and increases as $f(\infty)$ grows. Eq.(\ref{V16*}) is
now a straightforward consequence of the constraint (\ref{finf}) and
eq.(\ref{V18*}).

\section{Another argument for suppression of decay rate  at
small number of incoming particles}

In the previous section we have seen that
  the vacuum decay is always suppressed at small enough number of initial
particles. Now we  present
another argument for the suppression of the false vacuum decay in the limit
of a small number of incoming
 particles. Our discussion is based on
the arguments of ref.\cite{RT} that the rate of the decay
induced by a few particles can
be estimated by considering the maximum probability among all initial states
in the common eigenspace
 of the Hamiltonian and some other operator $\hat{A}$,
\[
  \hat{A}=\int dk A(k)a^{\dagger}_ka_k
\]
The usual choice is $A(k)=1$ for all
$k$, so the operator $\hat{A}$ is just
 the number of particles.
However, other choices of $A(k)$ are equally suitable, the only
restrictions being that $A(k)$ is non-negative at any $k$,
 does not depend explicitly on
the coupling constant ($v$ in our model) and does not grow like $|k|$
at large $k$ (otherwise $\hat{A}$ coincides with the Hamiltonian, up to
an unimporatnt piece; we assume for definiteness that $A(k)$ does not
grow at all at large momenta). If the eigenvalue $A$ of $\hat{A}$ is of
order $v^{2}$, the induced decay rate is semiclassically calculable
and gives the upper bound for the two-particle cross
 section\footnote{In fact, it has been conjectured
 in refs.\cite{RT,T} that
the leading exponent for the two-particle cross section can be
obtained by considering the limit $A/v^{2} \to 0$; we will not make
use of this conjecture in this paper.}.
So, we argue that the vacuum decay induced by a few highly energetic
particles is not exponentially suppressed only if there exist
real classical Minkowskian solutions with arbitrarily small $A/v^{2}$
which end up in the true vacuum, i.e., have a singularity
 $\phi = \infty$ at $x=0$ and finite $t$.

As usual, the eigenvalue of $\hat{A}$ in terms of the classical field
is
\[
    \hat{A}=\int dk \omega (k) A(k) \phi(k) \phi(-k)
\]
where $\phi$ is the asymptotics of the field at large negative time.
Let us choose
\[
  A(k) = \omega_k^{-1}
\]
We now show that at given energy $E$ there are no relevant classical solutions
 for
sufficiently small $A$. We again make use of the technique of
sect.2.1, and also use the notations of sect.3.
In complete analogy to eq.(\ref{nphi})
one has
\[
   A = {2\over\pi}\int\limits_0^{\infty}dk\phi_L(k)\phi_L(-k)
\]
We will minimize this expression for all functions $\phi_L$
under the condition that
the energy
\[
  E = {2\over\pi}\int\limits_0^{\infty}dk \omega^{2}_{k}\phi_L(k)\phi_L(-k)
\]
and $f(\infty)$ (\eq{finf}) are fixed. Introducing two Lagrange multipliers,
 $\Lambda^2$ and $\Lambda'^2$ we obtain the
following equation,
\begin{equation}
  \omega_k^2\phi_L(k) = \Lambda^2\phi_L(k)-\Lambda'^2\int dz\e^{-ikz}
  \exp\left({2\lambda\over v}\phi_L(z)\right)
  \label{momrepr}
\end{equation}
 For
further convenience let us introduce a new Lagrange parameter
 $\phi_0$ instead of
$\Lambda'$,
\[
  \Lambda'^2={\lambda\Lambda^2\phi_0^2\over v}
   \exp\left(-\frac{2\lambda}{v}\phi_{0}\right)
\]
Let us  assume that $\phi_0~\gtap~v$ and $\Lambda\gg m$; this assumption
will be justified a posteriori. Eq.(\ref{momrepr})
can be rewritten in the coordinate representation,
\[
  \partial_z^2\phi_L(z) = \Lambda^2\phi_L(z)-
  {\lambda\Lambda^2\phi_0^2\over v}
  \exp\left({2\lambda\over v}(\phi_L(z)-\phi_0)\right)
\]
The latter
 equation has the same form as the equation for the sphaleron in our
model. The solution in two different (and overlapping)
 regions  is as follows \cite{Rutgers}
\[
  \phi(z) = \phi_0 - {v\over\lambda}\ln\left[\cosh\left(
            {\lambda\Lambda\phi_0\over v}z\right)\right],
  ~~~\mbox{at } z\ll\Lambda^{-1}
\]
\[
  \phi(z) = \phi_0\e^{-\Lambda |z|},
            ~~~\mbox{at } z\gg (\lambda\Lambda)^{-1}
\]
For this field configuration, the value of $A$ is
\[
  A = {2\phi_0^2\over\Lambda}
\]
while the energy and $f(\infty)$ are
\begin{equation}
  E = 2\phi_0^2\Lambda
\label{13*}
\end{equation}
\begin{equation}
  f(\infty) = {2v\over\lambda\Lambda\phi_0}
              \exp\left({2\lambda\over v}\phi_0\right)
\label{13**}
\end{equation}
It is clear that the condition $f(\infty)>a$ imposes a lower bound on
 $A$
for a given value of energy $E$.  In the most interesting case of
 exponentially high energies, $\ln E/E_{sph} \sim \lambda$, one has
\[
  \frac{A}{v^{2}}>B(E)
\]
where
\[
   B(E)\sim \frac{E_{sph}}{mE}
\]
is strictly positive. Thus, at arbitrary but fixed energy, the
induced vacuum decay is classically forbidden in the limit
$A/v^{2}\to0$, i.e., in the limit of small number of incoming
particles.

Finally, let us justify the above assumptions on the Lagrange
parameters $\phi_{0}$ and $\Lambda$, again for simplicity
at exponentially high
energies. At $f(\infty)\sim a$ one finds from eqs.(\ref{13*}) and
(\ref{13**})
\[
  \Lambda \sim \frac{mE}{E_{sph}},~~~~~~~~
 \phi_{0}
 \sim \frac{v}{2}\left( 1 + \frac{1}{\lambda}\ln{\frac{E}{E_{sph}}}\right)
\]
so one indeed has $\Lambda \gg m$, $\phi_{0} \sim v$. This completes
the argument.

\section{Conclusions}

In this paper we have considered real classical solutions that
describe the unsuppressed vacuum decay. We have shown that for $n>n_{crit}$
these solutions exist when the energy is larger than some
 $n$--dependent
 critical
value. In the case  $\lambda^{-1}\ll n/n_{sph}\ll 1$, this critical
energy is essentially the same as
 the one found in our
previous study of the decay rate in the
suppressed regime \cite{Rutgers}.
 For $n<n_{crit}$, there are no such solutions,
so we  conclude that the decay rate is always suppressed if the number
of initial particles is small enough. We have presented
 an independent argument showing
that the induced vacuum decay is suppressed in the limit of small number of
initial particles. Our results,
 in particular, imply that the exponential
suppression of the decay of the false vacuum induced by two colliding
 particles persists at arbitrarily high energies. In other words, the
function $F(E)=v^{-2}\ln\Gamma(E)$, where $\Gamma(E)$ is the decay rate
induced by two particles with the
center-of-mass energy $E$, is always negative.

The absence of the classical Minkowskian solutions tells nothing
about the actual behavior of the suppression factor. The calculation
of $F(E)$ in our model
by the technique of the boundary value problem in complex
time also met technical problems at small $n$ and very high
energies \cite{Rutgers}. So, the (exponentially suppressed)
two-particle cross section of the
induced vacuum decay is not known in the interesting energy region.

There are two
possibilities of
how the function $F(E)$ may behave at high energies. The
first one
 is that $F$ tends to some finite value $F(\infty)>0$. In this case,
the exponential suppression is always strong (stronger than
$\e^{-v^2F(\infty)}$). The other possibility is that $F$ asymptotically
tends to zero at as $E\to\infty$. In that
 case, the induced false vacuum decay
may become "observable" at very high energies
(much higher than the one-instanton estimate), where the function $F$
actually becomes
small. Which of these possibilities is realized in our model is still an open
question.

The authors are indebted to T.~Banks and P.~Tinyakov for stimulating
disscussions. V.R. thanks Rutgers University, where part of this work
has been done, for hospitality.
 The work of D.T.S is supported in part by the Russian Foundation
for Fundamental Research (project 93-02-3812) and by the Weingart
Foundation through a cooperative agreement with the Department of Physics
at UCLA.

\newpage

FIGURE CAPTIONS

1. The potential $V(\phi)$.

2. The $(E,n)$ plane, $n_{crit}=\pi^{2}n_{sph}/\lambda$.
The regions II and I above and below the solid line correspond to
unsuppressed and suppressed induced vacuum decay, respectively. The
dashed line is the (unphysical) border between the regions of
suppressed and unsuppressed vacuum decay within the Ansatz of
 eq.(\ref{ansatz}).

\end{document}